\pdfoutput=1
\documentclass[usenatbib]{mn2e}
\usepackage{amsfonts,amsmath,amssymb,amsbsy}
\usepackage{apjfonts}
\usepackage{epsfig}
\usepackage{fixltx2e} 
\usepackage{verbatim}
\usepackage{url}
\usepackage[colorlinks,citecolor=blue]{hyperref}

\newcommand{\be}{\begin{equation}}
\newcommand{\ee}{\end{equation}}

\newcommand{\GIZMO}{{\small GIZMO}}
\newcommand{\gizmourl}{\href{http://www.tapir.caltech.edu/~phopkins/Site/GIZMO.html}{\nolinkurl{www.tapir.caltech.edu/~phopkins/Site/GIZMO}}}
\newcommand{\gizmopublicurl}{\href{https://bitbucket.org/phopkins/gizmo-public}{\nolinkurl{https://bitbucket.org/phopkins/gizmo-public}}}
\newcommand{\UserGuide}{\href{http://www.tapir.caltech.edu/~phopkins/Site/GIZMO_files/gizmo_documentation.html}{\em User Guide}}

\newcommand{\plotsidesize}[2]{\centering \leavevmode \includegraphics[width={#2\textwidth}]{#1}}

\newcommand\altaffiltext[1]{$^{#1}$}
\title[GIZMO 2017]{A New Public Release of the GIZMO Code\vspace{-0.5cm}}

\vspace{-0.2cm}
\author[P.~F.\ Hopkins]{
\parbox[t]{\textwidth}{ 
Philip F.~Hopkins
} 
\vspace*{6pt} \\
\altaffiltext{}{Theoretical Astrophysics (TAPIR), Mailcode 350-17, California Institute of Technology, Pasadena, CA 91125, USA. Email:phopkins@caltech.edu}
\vspace{-0.9cm}
}

\date{\ December, 2017\vspace{-0.6cm}}
\begin{document}
\maketitle
\label{firstpage}

\vspace{-0.5cm}
\begin{abstract}
\vspace{-0.1cm}

We describe a major update to the public \GIZMO\ code. \GIZMO\ has been used in simulations of cosmology; galaxy and star formation and evolution; black hole accretion and feedback; proto-stellar disk dynamics and planet formation; fluid dynamics and plasma physics; dust-gas dynamics; giant impacts and solid-body interactions; collisionless gravitational dynamics; and more. This release of the public code supports: hydrodynamics (using various mesh-free finite-volume Godunov methods or SPH), ideal and non-ideal MHD, anisotropic conduction and viscosity, radiative cooling and chemistry, star and black hole formation and feedback, sink particles, dust-gas (aero)-dynamics (with or without magnetic fields), elastic/plastic dynamics, arbitrary (gas, stellar, degenerate, solid/liquid material) equations of state, passive scalar/turbulent diffusion, large-eddy and shearing boxes, self-gravity with fully-adaptive force softenings, arbitrary cosmological expansion, and on-the-fly group-finding. It is massively-parallel with hybrid MPI+OpenMP scaling verified up to $>1$ million threads. The code is extensively documented, with test problems and tutorials provided for these different physics modules. 

\end{abstract}

\begin{keywords}
methods: numerical --- hydrodynamics --- instabilities --- turbulence --- planets: formation --- stars: formation --- galaxies: formation --- cosmology: theory
\vspace{-1.0cm}
\end{keywords}

\vspace{-1.1cm}
\section{Overview}
\label{sec:intro}

This is an announcement and very brief description of the new public release of the \GIZMO\ code, available at:\\
\gizmourl\\
or:\\
\gizmopublicurl\\

The simulation code \GIZMO\ \citep{hopkins:gizmo} is a flexible, arbitrary Lagrangian-Eulerian multi-method code, with a wide range of different physics modules described below. The code is descended from {\small P-GADGET} and {\small GADGET-2} \citep{springel:gadget}, and is fully-compatible with {\small GADGET} analysis codes, snapshots, and initial conditions. 

{\bf Documentation:} \GIZMO\ is extensively documented. Users should read the \UserGuide, bundled with the code. This contains detailed descriptions of all code modules, compiler flags, run-time parameters, initial conditions files, snapshots, etc., as well as instructions for building or analyzing your own simulations (and links to many public codes which can generate initial conditions or visualize outputs). Dozens of test problems with complete tutorials, setup, and initial conditions are provided in the \UserGuide.

{\bf Compatibility:} \GIZMO\ is written in standard C, and uses only widely-available public libraries. The portability of the code has been confirmed on a large number of systems, ranging from NSF's Comet, Stampede, and Blue Waters, NASA's Pleiades, and DOE's Titan and Mira super-clusters, through Mac and Linux laptops.

{\bf Scaling and Parallelism:} \GIZMO\ is a massively-parallel code which uses a hybrid OpenMP+MPI architecture to efficiently scale to large numbers of CPU cores and/or threads. While code scaling is always highly problem-dependent, actual production problems (e.g.\ large cosmological hydrodynamic simulations with star formation and magnetic fields) have achieved near-ideal strong or weak scalings through $>1$ million threads. 

{\bf Development Code:} Some modules -- for example radiation-hydrodynamics -- are not yet in the public code because they are in active development and not yet de-bugged or tested at the level required for use ``out of the box'' (but will be made public as soon as this stage is reached). Other modules (e.g.\ the {\small FIRE} project feedback physics) are not public because they involve proprietary code developed by others with their own collaboration policies, which must be respected (see \UserGuide\ for details). 

{\bf License:} The public version of the code is free software, distributed under the GNU General Public License. Read the \UserGuide\ for more details. The authors retain their copyright on the code. The development version of the code is private and can only be accessed or shared with the explicit permission of the authors.

\begin{figure*}
\plotsidesize{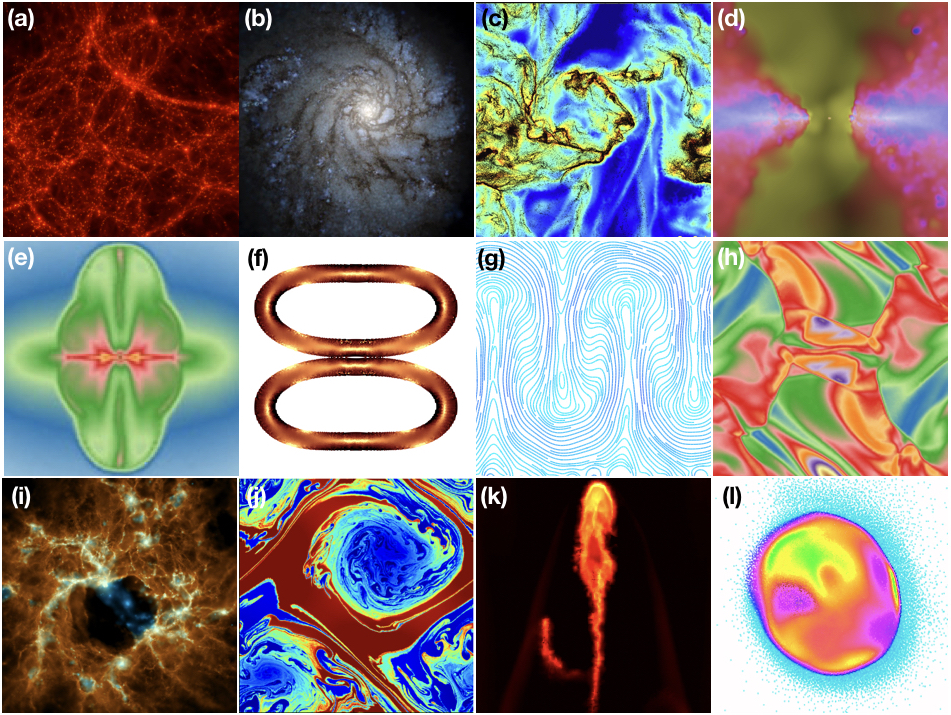}{0.99}
    \vspace{-0.25cm}
    \caption{Examples of simulations run with \GIZMO. {\bf (a)} Cosmology: Large-volume dark matter+baryonic cosmological simulation \citep{dave.2016:mufasa.fire.inspired.cosmo.boxes}. {\bf (b)} Galaxy formation: with cooling, star formation, and stellar feedback \citep{wetzel.2016:latte}. {\bf (c)} Dust (aero)-dynamics: grains moving with Epstein drag and Lorentz forces in a super-sonically turbulent cloud \citep{lee:dynamics.charged.dust.gmcs}. {\bf (d)} Black holes: AGN accretion according to resolved gravitational capture, with feedback driving outflows from the inner disk \citep{hopkins:qso.stellar.fb.together}. {\bf (e)} Proto-stellar disks: magnetic jet formation in a symmetric disk around a sink particle, in non-ideal MHD \citep{raives:communication}. {\bf (f)} Elastic/solid-body dynamics: collision and compression of two rubber rings with elastic stresses and von Mises stress yields. {\bf (g)} Plasma physics: the magneto-thermal instability in a stratified plasma in the kinetic MHD limit \citep{hopkins:gizmo.diffusion}. {\bf (h)} Magneto-hydrodynamics: the Orszag-Tang vortex as a test of strongly-magnetized turbulence \citep{hopkins:mhd.gizmo}. {\bf (i)} Star formation: star cluster formation (with individual stars) including stellar evolution, mass loss, radiation, and collisionless dynamics \citep{grudic:sfe.cluster.form.surface.density}. {\bf (j)} Fluid dynamics: the non-linear Kelvin-Helmholtz instability \citep{hopkins:gizmo}. {\bf (k)} Multi-phase fluids in the ISM/CGM/IGM: ablation of a cometary cloud in a hot blastwave with anisotropic conduction and viscosity \citep{su:2016.weak.mhd.cond.visc.turbdiff.fx}. {\bf (l)} Impact and multi-material simulations: giant impact (lunar formation) simulation, after impact, showing mixing of different materials using a Tillotson equation-of-state \citep{deng:giant.impact.sim.mfm}.
    \vspace{-0.25cm}
        \label{fig:demo}}
\end{figure*}

\begin{figure*}
\centering
\includegraphics[width = 0.33 \textwidth]{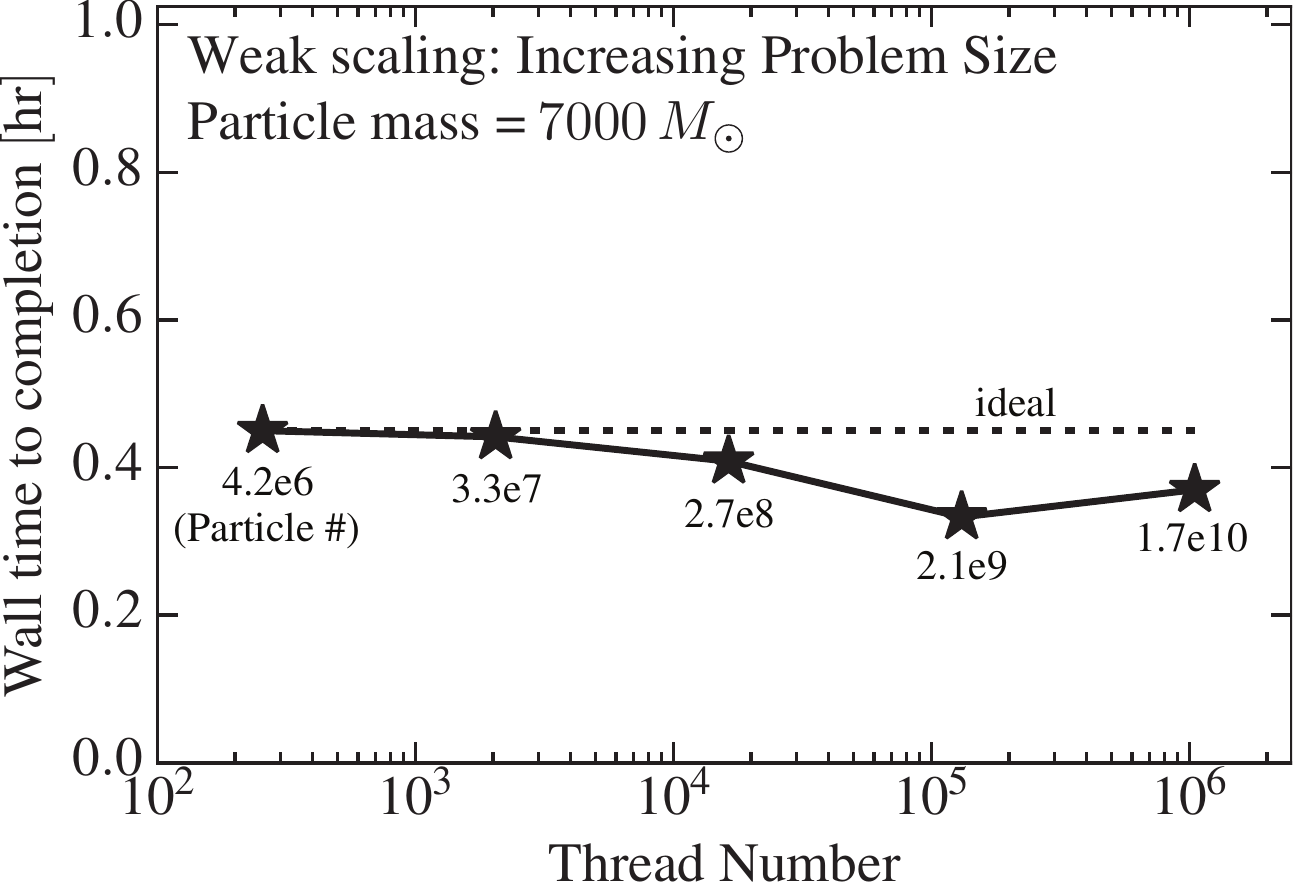} 
\includegraphics[width = 0.33 \textwidth]{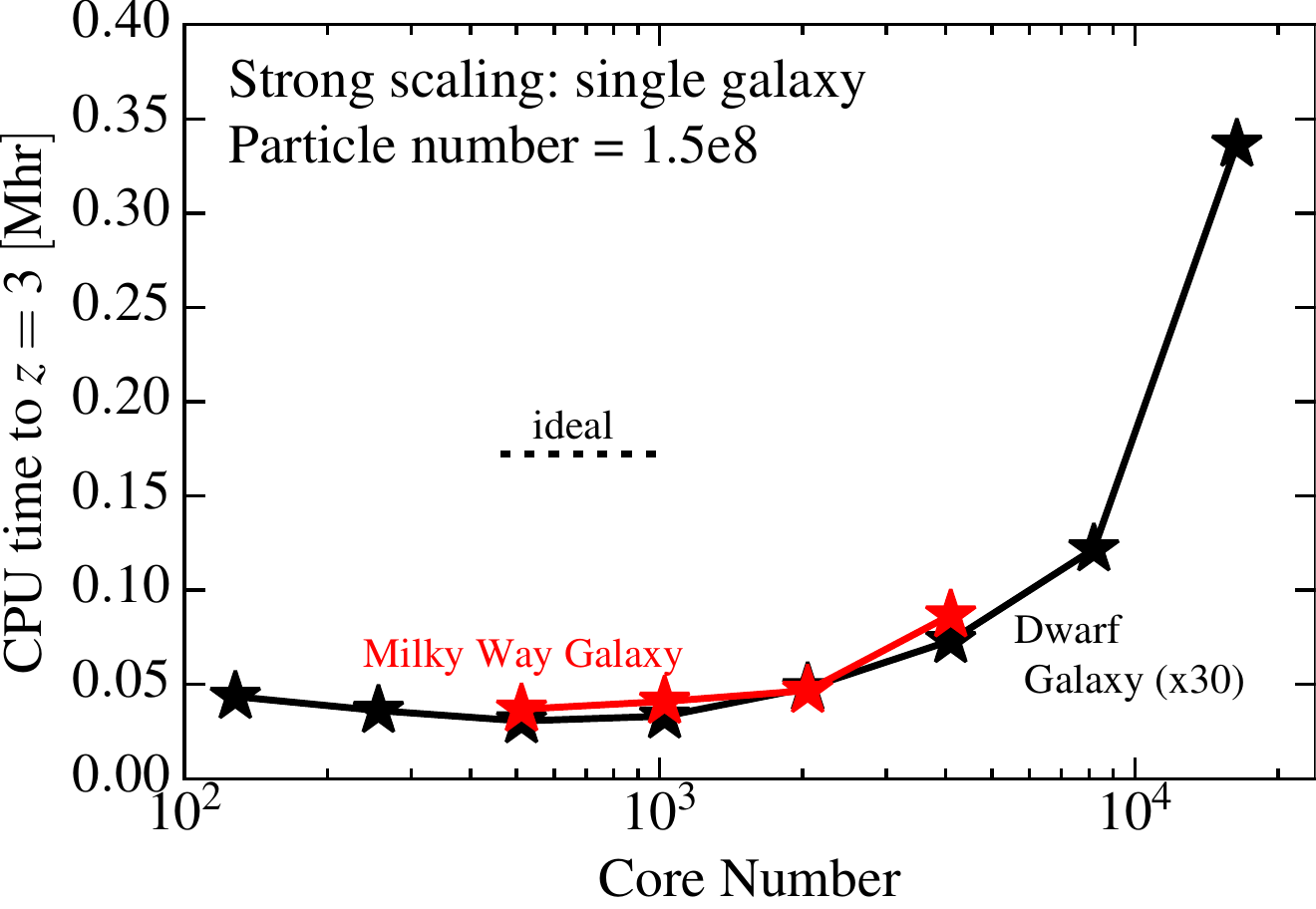} 
\includegraphics[width = 0.33 \textwidth]{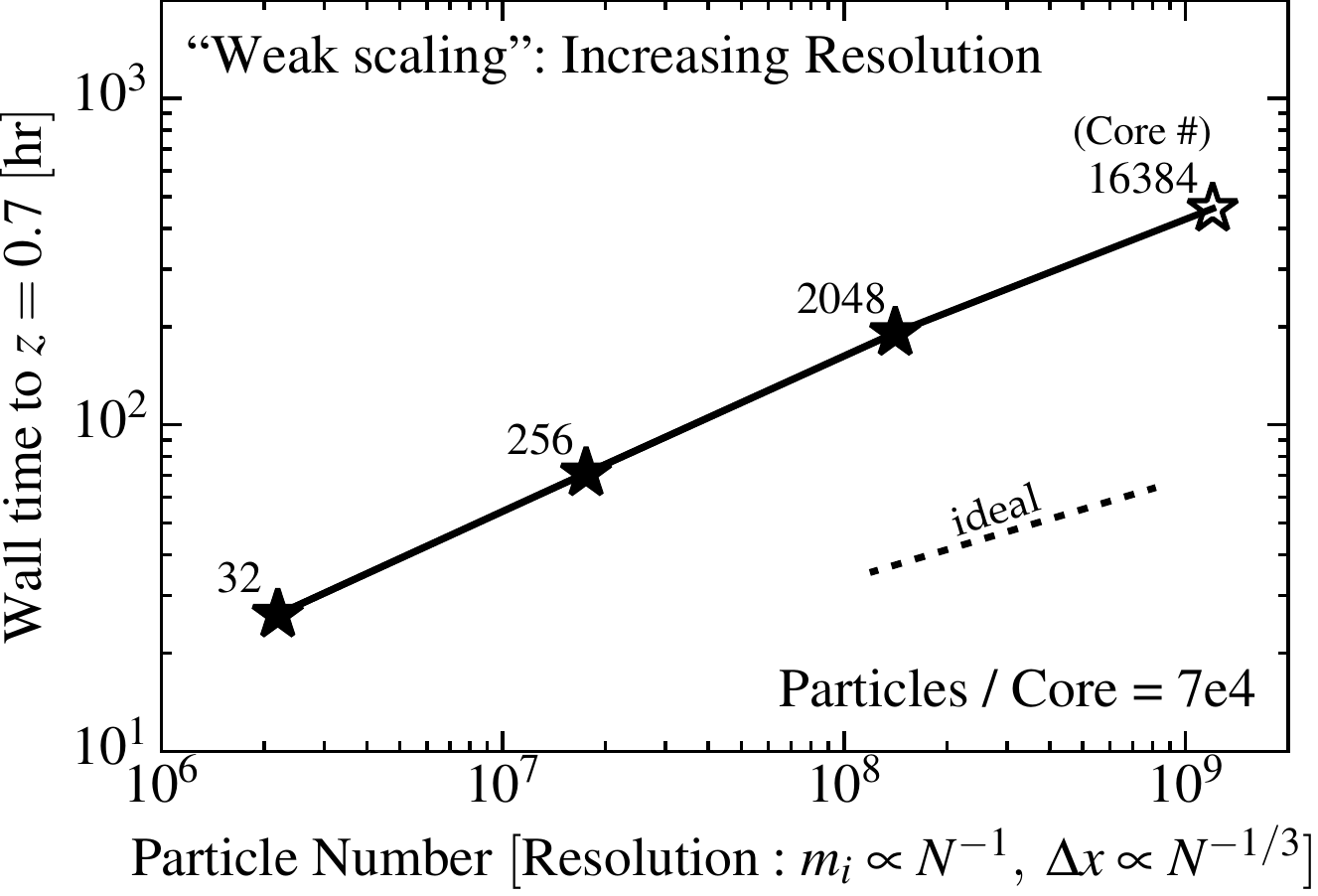}
\caption{
Code scalings of \GIZMO\ in cosmological simulations of galaxy formation, including self-gravity, hydrodynamics, cooling, star formation, and stellar feedback \citep{hopkins:fire2.methods}. 
{\em Left:} Weak scaling (making the computational problem larger in size, but increasing the processor number proportionally) for a full cosmological box, populated with high resolution particles (of fixed mass), run for a short fraction of the age of the Universe. We increase the cosmological volume from $2$ to $10^{4}\,{\rm Mpc^{3}}$. Theoretically ``ideal'' weak scaling would be flat (dashed line). The weak scaling of \GIZMO\ is near-ideal (actually slightly better at intermediate volume, owing to fixed overheads and statistical homogeneity of the volume at larger sizes) to greater than a million threads ($>250,000$ CPUs and $>16,000$ nodes). 
{\em Center:} Strong scaling (fixed-size problem, increasing the processor number) for a ``zoom-in'' simulation of a Milky Way-mass galaxy ($10^{12}\,M_{\sun}$ dark matter halo mass, $10^{11}\,M_{\sun}$ stellar mass) or a dwarf galaxy ($10^{10}\,M_{\sun}$ halo, $10^{6}\,M_{\sun}$ stellar), each using 150 million particles in the galaxy. Our optimizations allow us to maintain near-ideal strong scaling (flat, parallel to dashed line) on this problem up to $\sim16,000$ cores per billion particles ($\sim 2000$ at the specific resolution shown). 
{\em Right:} Increasing the resolution instead of the problem size (specifically, increasing the particle number for the same Milky Way-mass galaxy, while increasing the processor number proportionally). Because the resolution increases, the time-step decreases (smaller structures are resolved), so the ideal weak scaling increases following the dashed line. The achieved scaling is very close to ideal, up to $>16,000$ cores for $\sim1$ billion particles. This is a highly inhomogeneous, high-dynamic-range (hence computationally challenging) problem with very dense structures (e.g. star clusters) taking very small timesteps ($<100\,$yr), while other regions (e.g. the inter-galactic medium) take large timesteps (up to $\sim 10^{7}\,$yr) -- more ``uniform'' problems (e.g.\ dark-matter-only cosmological or pure-hydrodynamic turbulence simulations) can achieve even better scalings. 
\label{fig:scaling}
}
\vspace{-0.5cm}
\end{figure*}

\vspace{-0.6cm}
\section{Physics in the Public Code}

\GIZMO\ is a modular, multi-physics code; some examples of different \GIZMO\ simulations are shown in Fig.~\ref{fig:demo}. This public release adds support for a range of physics including (but not limited to): 

\begin{itemize}

\item {\bf Hydrodynamics} using any of several fundamentally different methods (e.g.\ new Lagrangian finite-element Godunov schemes, or various ``flavors'' of smoothed particle hydrodynamics, or Eulerian fixed-grid schemes). The fluid ``mesh'' can be assigned any arbitrary motion the user desires (or allowed to move with the fluid). 

\item {\bf Ideal and non-ideal magneto-hydrodynamics (MHD)}, including Ohmic resistivity, ambipolar diffusion, and the Hall effect. Coefficients can be set by-hand or calculated self-consistently from the code chemistry.

\item {\bf Radiative heating and cooling}, including pre-built libraries with photo-ionization, photo-electric, dust, Compton, Brehmstrahhlung, recombination, fine structure, molecular, species-by-species metal line cooling, or hooks for popular external chemistry and cooling libraries. 

\item {\bf Star formation \&\ feedback} on galactic scales (for galaxy-formation or ISM studies) or individual-star scales (for IMF studies), including conversion of gas into sink particles according to various user-defined criteria (e.g.\ density, self-gravity, molecular thresholds). This includes mass return and feedback with resolved mass, metal, and energy input into the surrounding ISM, or various popular sub-grid galactic wind models. 

\item {\bf Black holes} including on-the-fly formation and seeding with user-defined criteria (based on local or group properties), mergers, growth via various sub-grid accretion models (with or without models for an un-resolved accretion disk), or via explicitly resolved gravitational capture, and ``feedback'' scaled with the accretion.

\item {\bf Elastic \&\ plastic dynamics}, with support for arbitrary Tillotson-type equations-of-state for solid, liquid, or vapor media, negative-pressure media, anisotropic deviatoric stresses and plasticity, and various pre-defined material properties.

\item {\bf Arbitrary equations-of-state} (EOSs), including support for trivially-specifiable modular EOSs, multi-fluid systems, and pre-programmed EOSs for stellar systems or degenerate objects (Helmholtz-type EOSs), as well as solid/liquid/vapor mixtures (Tillotson-type EOSs). 

\item {\bf Sink particles}, with dynamical accretion and formation properties, and user-specified formation conditions from local gas or group properties. 

\item {\bf Anisotropic conduction and viscosity}: ``real'' Navier-Stokes or fully-anisotropic Spitzer-Braginskii conduction and viscosity, with dynamically calculated or arbitrarily chosen coefficients. 

\item {\bf Dust-gas mixtures}, e.g.\ aerodynamically coupled grains or other particles. The treatment is flexible and can handle both sub and super-sonic cases, compressible fluids, and grain-gas back-reaction, with arbitrary dust drag laws (Epstein, Stokes, Coulomb drag) and Lorentz forces on charged grains from magnetic fields.

\item {\bf Self-gravity} for arbitrary geometries and boundary conditions, with fully-adaptive Lagrangian gravitational softenings for all fluids and particle types. Arbitrary analytic potentials can be added trivially.

\item {\bf Turbulent eddy diffusion} of passive scalars and dynamical quantities (Smagorinski diffusion for subgrid-scale turbulence).

\item {\bf Shearing boxes} (stratified or unstratified), {\bf ``large-eddy simulations''} (driven turbulent boxes), periodic, open, or reflecting boundary conditions are supported.

\item {\bf Cosmological integrations} (both large-volume and ``zoom-in'' cosmological simulations), with support for ``non-standard'' cosmologies including dynamical dark energy equations-of-state or arbitrarily time-dependent expansion histories or gravitational constants.

\item {\bf Group-finding} and/or power-spectrum computation, run on-the-fly, for arbitrary combinations of ``target'' species (e.g.\ halo or galaxy or cluster-finding).

\item {\bf Particle splitting/merging} according to arbitrary user-defined criteria, to allow for super-Lagrangian ``hyper-refinement''  simulations.

\end{itemize}

\vspace{-0.8cm}
\section{Code Scaling \&\ Performance}

\GIZMO\ employs a hybrid MPI+OpenMP parallelization strategy with a flexible domain decomposition and hierarchical adaptive timesteps (together with a large number of optimizations for different problems and physics), which enable it to scale efficiently on massively-parallel systems with problem sizes up to and beyond billions of resolution elements \citep[][]{hopkins:fire2.methods}. 
Code scalings are always (highly) problem-and-resolution-dependent, but illustrative examples of scalings for real ``production'' simulations (run on the DOE Titan and Mira clusters) are shown in Fig.~\ref{fig:scaling}.

\vspace{-0.5cm}
\section{Arbitrary (Moving) Meshes}

\GIZMO\ is multi-method in that the ``mesh'' over which the fluid equations is solved can be specified with arbitrary motion or geometry (or lack thereof). \GIZMO\ can be run as a Lagrangian mesh-free finite-volume code (where the mesh moves with the fluid), or as an SPH code, or as a fixed Cartesian grid code (similar to codes like {\small ATHENA} and {\small ZEUS}). More generally, users can specify any background mesh motion they desire (e.g.\ shearing, expanding, collapsing, accelerating, or differentially rotating boxes).

\vspace{-0.5cm}
\section{Thanks \&\ Acknowledgments}

We thank Volker Springel both for his personal mentoring, and for writing {\small GADGET}, without which \GIZMO\ would not exist. We also thank the large number of \GIZMO\ code developers who have contributed to the content in the public code, especially Daniel Angles-Alcazar, Xiangcheng Ma, Shea Garrison-Kimmel, Mike Grudic, Alessandro Lupi, Robert Thompson, Quirong Zhu, Hongping Deng, Dusan Keres, and Paul Torrey. 

Support for PFH and \GIZMO\ development was provided by an Alfred P. Sloan Research Fellowship, NSF Collaborative Research Grant \#1715847 and CAREER grant \#1455342, Caltech compute cluster ``Wheeler,'' allocations from XSEDE TG-AST130039 and PRAC NSF.1713353 supported by the NSF, and NASA HEC SMD-16-7592.




\vspace{-0.5cm}
\bibliography{ms_local}


\end{document}